\documentclass{llncs}

\usepackage[pdftex]{graphicx}
\usepackage[caption=false,font=footnotesize]{subfig}

\graphicspath{{images/}}

\usepackage{xcolor}
\usepackage{svg}
\usepackage[hyphens]{url}
\usepackage{footnote}
\usepackage{tikz}
\usepackage{tikzscale}
\usepackage{pgfplots}
\usepackage{wrapfig}
\usetikzlibrary{decorations.pathreplacing,patterns}
\makesavenoteenv{tabular}
\usepackage[font=scriptsize]{caption}
\usepackage{hyperref}
\usepackage{float}
\usepackage{algorithm}
\usepackage[noend]{algpseudocode}
\algrenewcommand\alglinenumber[1]{\scriptsize #1:}
\usepackage{bbding}

\begin{document}

\title{Disaggregating Non-Volatile Memory for Throughput-Oriented Genomics Workloads}

\author{Aaron Call\inst{1,2}\Envelope, Jord\`{a} Polo\inst{1}, David Carrera\inst{1,2}, Francesc Guim\inst{3}, Sujoy Sen\inst{3}}
\institute{Barcelona Supercomputing Center (BSC)
\email{\{aaron.call,jorda.polo,david.carrera\}@bsc.es}
\and Universitat Polit\`ecnica de Catalunya (UPC)
\and Intel Corporation\\
\email{\{francesc.guim,sujoy.sen\}@intel.com}
}

\maketitle

\begin{abstract}
    Massive exploitation of next-generation sequencing technologies requires
    dealing with both: huge amounts of data and complex bioinformatics
    pipelines. Computing architectures have evolved to deal with these
    problems, enabling approaches that were unfeasible years ago: accelerators
    and Non-Volatile Memories (NVM) are becoming widely used to enhance the
    most demanding workloads. However, bioinformatics workloads are usually
    part of bigger pipelines with different and dynamic needs in terms of
    resources. The introduction of Software Defined Infrastructures (SDI) for
    data centers provides roots to dramatically increase the efficiency in the
    management of infrastructures. SDI enables new ways to structure hardware
    resources through disaggregation, and provides new hardware composability
    and sharing mechanisms to deploy workloads in more flexible ways.  In this
    paper we study a state-of-the-art genomics application, SMUFIN, aiming to
    address the challenges of future HPC facilities.

\end{abstract}


\section{Introduction}

The genetic basis of disease is increasingly becoming more accessible thanks
to the emergence of Next Generation Sequencing platforms, which have extremely
reduced the costs and increased the throughput of genomic sequencing. For the
first time in history, personalized medicine is close to becoming a reality
through the analysis of each patient's genome. Genomic variations, between
patients or among cells of the same patient, have been identified to be the
direct cause, or a predisposition to genetic diseases: from single nucleotide
variants to structural variants, which can correspond to deletions,
insertions, inversions, translocations and copy number variations, ranging
from a few nucleotides to large genomic regions.

The exploitation of genomic sequencing should involve the accurate
identification of all kinds of variants, in order to derive a correct
diagnosis and to select the best therapy. For clinical purposes, it is
important that this computational process be carried out within an effective
timeframe. But a simple sequencing experiment typically yields thousands of
millions of reads per genome, which have to be stored and processed. As a
consequence, the analysis of genomes with diagnostic and therapeutic purposes
is still a great challenge, both in the design of efficient algorithms and at
the level of computing performance.

The field of computational genomics is quickly evolving in a continuous seek
for more accurate results, but also looking for improvements in terms
of performance and cost-efficiency. In parallel, computing architectures have
also evolved, enabling approaches that were unfeasible only years ago. The
use of Non-Volatile Memories (NVM) and accelerators has been widely adopted for
all kinds of workloads with the introduction of NVMe cards, GPUs, and FPGAs
for some of the most demanding computing challenges. Genomics workloads today
have a larger variety of requirements related to
the compute platforms they run in. Workloads are tuned to work optimally on
specific configurations of compute, memory, and storage. On top of that,
current genomics workloads and pipelines tend to be composed of multiple
phases with different behaviors and resource requirements.

One such example in the context of variant calling is SMUFIN~\cite{smufin1}, a
state-of-the-art method that performs a direct comparison of normal and tumor
genomic samples from the same patient without the need of a reference genome,
leading to more comprehensive results. In its original implementation,
published in Nature~\cite{smufin1} in 2014, this novel approach required
significant amounts of resources in a supercomputing facility. Since then, it
has been optimized and adapted to scale up and make the most of Non-Volatile
Memory~\cite{smufin2scaleup}.

Beyond Non-Volatile Memories and accelerators, new technological advances
currently under development, such as Software Defined Infrastructures, are
dramatically changing the data center landscape.  One of the key features of
Software Defined Infrastructures is \textit{disaggregation}, which allows
dynamically attaching and detaching resources from physical nodes with just a
software operation, removing the constraints of getting hardware components
statically confined to servers.
This paper takes a modern genomics workload, SMUFIN, evaluates disaggregation
mechanisms when running it, and describes how characterization can be used to
guide the orchestration of a genomics pipeline.

The rest of the paper is structured as follows. Section~\ref{s_smufin}
provides an overview of the foundations of SMUFIN, the variant-calling method
studied in this paper.  Section~\ref{s_disaggregation} introduces resource
disaggregation and the technology used to implement it.  Next,
Section~\ref{sec:evaluation} characterizes disaggregation mechanisms using
SMUFIN.  Section~\ref{s_orchestration} shows how characterization can be used
to guide orchestration.  And finally, Section~\ref{sec:related} discussed
related work and Section~\ref{sec:conclusions} concludes.

\section{SMUFIN: A Throughput-oriented Genomics Workload}
\label{s_smufin}
Most currently available methods for detecting genomic variations rely on an
initial step that involves aligning sequence reads to a reference genome
generally using Burrows-Wheeler transform~\cite{bwt}, which has an impact not
only on performance, but also on the accuracy of results. First, tumoral reads
that carry variation may be harder or impossible to align against a reference
genome. Second, the use of references also leads to interference with millions
of inherited (germline) variants that affect the actual identification of
somatic changes,
consequently decreasing the
final reliability and applicability of the results.
The initial alignment also has an impact on
subsequent analysis since most methods are tuned to identify only a particular
kind or size of mutation~\cite{medvedev}.
Alternative methods that don't rely on the initial alignment of sequenced reads
against a reference genome have been developed. In particular, the application
used in this work is based on SMUFIN~\cite{smufin1}, a reference-free approach
based on a direct comparison between normal and tumoral samples from the same
patient. The basic idea behind SMUFIN can be summarized in the following
steps: (i) input two sets of nucleic acid reads, normal and tumoral; (ii) build
frequency counters of substrings in the input reads; and (iii) compare branches
to find imbalances, which are then extracted as candidate positions for
variation.

\begin{figure}[t]
\vspace{-2.0em}
\centering
\includegraphics[width=0.75\columnwidth]{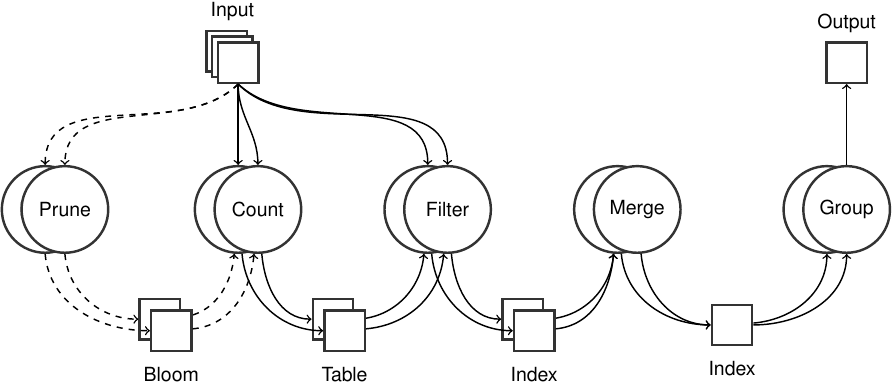}
\caption{SMUFIN's variant calling architecture: overview of stages and its data flow}
\label{f_overview}
\vspace{-2.0em}
\end{figure}

Internally, SMUFIN consists of a set of checkpointable stages
that are combined to build fully fledged workloads (Figure~\ref{f_overview}).
These stages can be shaped on computing platforms depending on different
criteria, such as availability or cost-effectiveness, allowing executions to
be adapted to its environment. Data can be split into one
or more \textit{partitions}, and each one of these partitions can then be
placed and distributed as needed: sequentially in a single machine, in
parallel in multiple nodes, or even in different hardware depending on the
characteristics of the stage.
Data partitioning can be effectively used to adapt executions to a particular
level of resources made available to SMUFIN, because it imposes a trade-off
between computation and IO. This data partitioning can be achieved by going
multiple times through the input data set that corresponds to each stage:
\textit{Prune}, \textit{Count}, and \textit{Filter}. In practice, systems with
high-end capabilities will not require a high level of partitioning and hence
IO, what ends up with scale-up solutions; on the opposite side of the
spectrum, lower-end platforms are able to run the algorithm by partitioning
data and duplicating IO, leading to scale-out solutions.  The goal of each one
of the stages is as follows:

\begin{itemize}

    \item \textit{Prune}: Discards sequences from the input by generating a
        bloom filter of k-mers that have been observed in the input more than
        once. Allows lowering memory requirements at the expense of additional
        computation and IO.

    \item \textit{Count}: Builds a frequency table of normal and tumoral
        k-mers in the input sequences.
        More specifically, k-mer counters are used to detect imbalances when
        comparing two samples.

    \item \textit{Filter}: Selects k-mers with imbalanced frequencies, which
        are candidates for variation, while also building 
        indexes of sequences with such k-mers.

    \item \textit{Merge}: Reads and combines multiple filter indexes from
        different partitions into single, unified indexes. Merging indexes
        only involves simple operations such as concatenation, OR on bitmaps,
        and appending.

    \item \textit{Group}: Matches candidate sequences that belong to the same
        region. First, selecting reads that meet certain
        criteria,
        and then retrieving related reads by looking up those that contain the same
        imbalanced k-mers.

\end{itemize}

One of the main characteristics of the current version of
SMUFIN~\cite{smufin2scaleup} is its ability to use NVM as memory extension.
This can be exploited in two different ways. First, using an NVM optimized
Key-Value Store such as RocksDB, and second, using a custom optimized swapping
mechanism to flush memory directly to the device.
When such memory extensions are available, a maximum size for the data
structures is set; once such size is reached, data is flushed to the memory
extension while a new empty structure becomes available. Generally speaking, bigger sizes are recommended: they help avoid
duplicate data, and also lead to higher performance, as writing big chunks to
a Non-Volatile Memory allows to exploit internal parallelism typical of flash
drives~\cite{chen_essential_2011}.

\begin{figure}
\vspace{-1.0em}
\centering
\includegraphics[width=0.60\columnwidth]{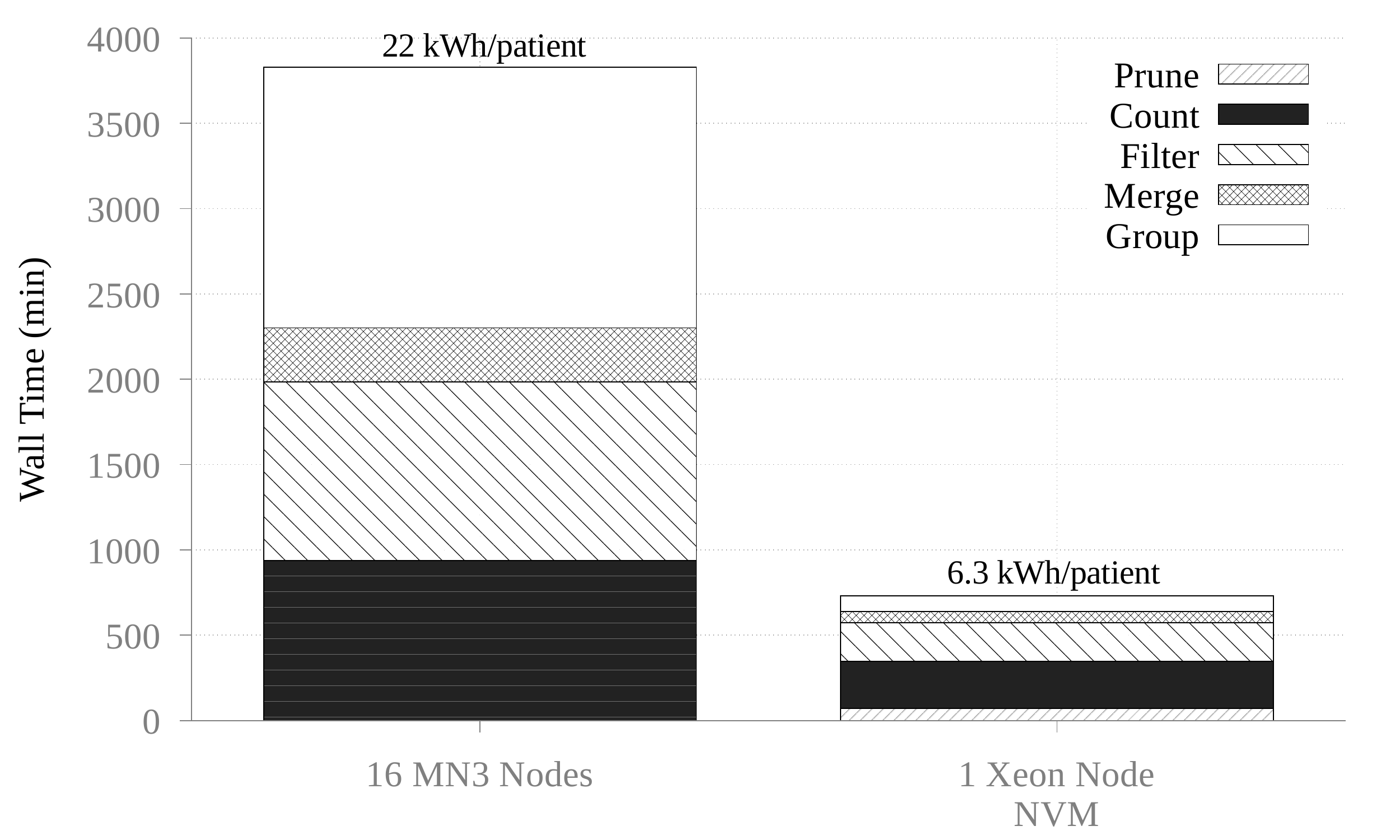}
\vspace{-0.5em}
\caption{Aggregate CPU time of a SMUFIN execution
    running in 16 MareNostrum nodes and in 1 Xeon-based node with NVM. Power
    consumption per execution (one patient) shown for reference.}
\label{f_smufin_time}
\vspace{-2.2em}
\end{figure}

SMUFIN's performance greatly benefits from NVM, as shown in
Figure~\ref{f_smufin_time}, which compares an execution in 16 machines in a
supercomputing facility (left) and a scale-up execution in a single node with NVM
enabled (right). The latter leads to faster executions and
lower power consumption. NVM can be leveraged in some way in most SMUFIN stages, and the
experiments performed in this paper are focused on \textit{Merge} using the
RocksDB-based implementation, which is one of the most IO intensive of the
pipeline.  However, other stages have similar characteristics and the same
techniques can be used elsewhere.

\section{Resource Disaggregation}
\label{s_disaggregation}

Traditional data centers usually contain homogeneous and heterogeneous compute
platforms (also referred to as computing nodes or servers). These platforms
are statically composed by computing, memory, storage, fabric, and/or
accelerator components, and they are usually stored in racks. However, in the
last few years there has been a trend towards new technologies that allow
disaggregating resources over the network, increasing flexibility and easying
the management of such data centers.

This paper analyzes the use of one of those new technologies: \textit{NVMe
Over Fabrics} (NVMeOF). First off, NVMe~\cite{nvme} is an interface
specification for accessing direct-attached NVM via a regular
PCI Express bus. On the other hand NVMeOF~\cite{nvmeof} is an emerging network protocol used to communicate
nodes with NVMe devices over a networking fabric. The architecture of NVMeOF
allows scaling to large numbers of devices, and supports a range of
different network fabrics, usually through Remote Direct Memory Access (RDMA)
so as to eliminate middle software layers and provide very low latency.

Disaggregating NVMe over the network with NVMeOF allows new mechanisms to
scale-up and improve efficiency of genomics workloads:

\begin{description}

    \item[Resource Sharing.] As workloads perceive remote NVMe as
        physically attached to their compute nodes, those can be partitioned,
        and each one of these partitions can then be exposed to the
        computational nodes as an exclusive resource. This translates into
        workload-unaware resource sharing, which in turn can lead to improved
        resource efficiency by maximizing usage.

    \item[Resource Composition.] Certain resources can be aggregated
        and exposed as a single, physically attached resource.  Instead of
        accessing individual units, accessing combined resources enables
        increased capacities that can lead to improved performance.  For
        instance, two SSD disks with a bandwidth of 2GB/s each can be composed
        and exposed as a single one with twice as much capacity and bandwidth,
        providing a total of 4GB/s.

\end{description}

\section{Characterizing Resource Disaggregation on SMUFIN}
\label{sec:evaluation}

\begin{figure*}[t]
	\centering
	\includegraphics[width=1\textwidth]{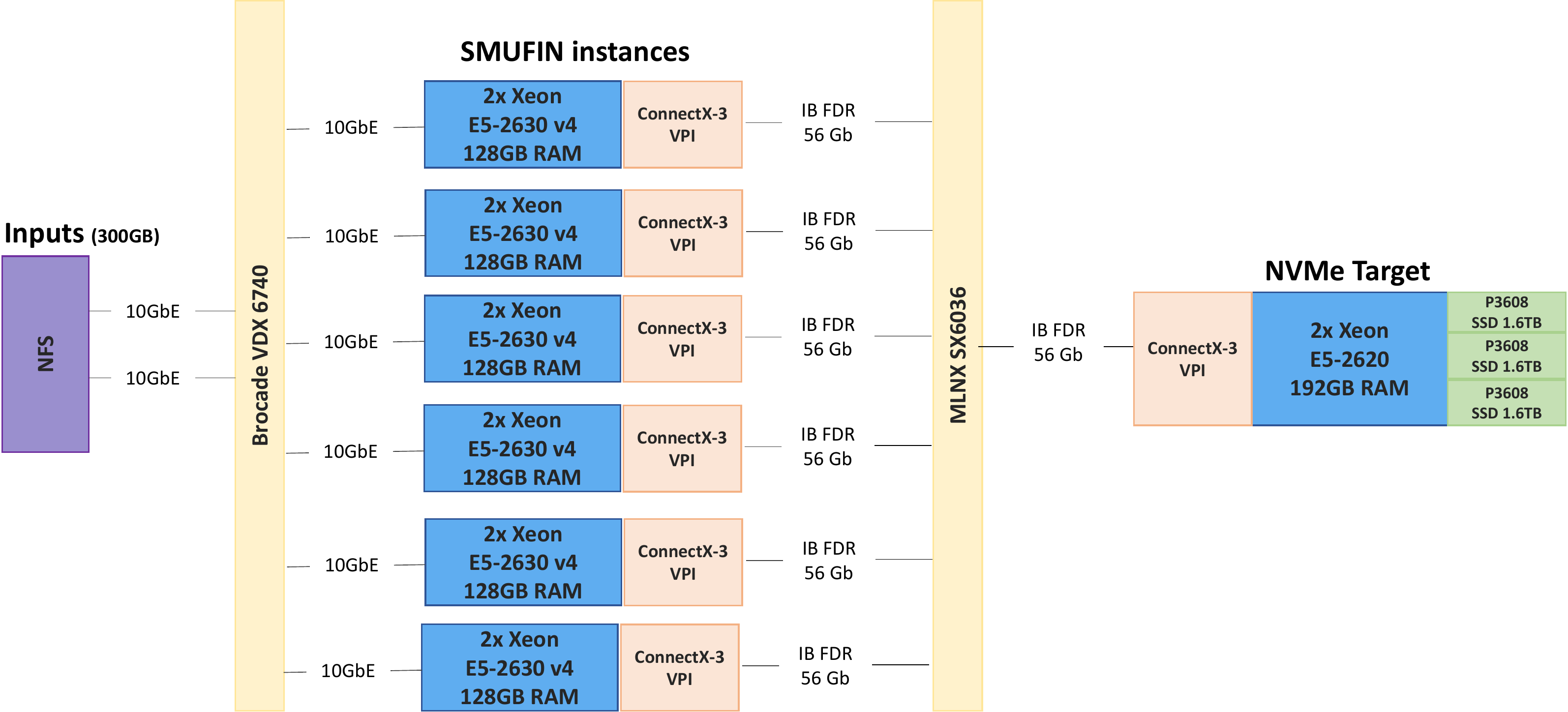}
	\caption{Experiments environment}
	\label{fig:exp_env}
\end{figure*}

In a continuous need to deal with increasingly larger amounts of data,
genomics workloads are quickly adapting, and NVM
technologies have become widely used as a key component in the memory-storage
hierarchy. This Section explores how disaggregating NVM might have an impact
on genomics workloads, and in particular SMUFIN. As part of the evaluation,
resource sharing and composition are analyzed using NVMeOF in an attempt to
scale-up and shape the performance of the workload.

\subsection{Experimental Environment}

The experiments are conducted in an environment as depicted in figure
\ref{fig:exp_env}.

The NVMe drives are used by SMUFIN as a memory
extension over fabric to store temporary data structures required to
accelerate the computation. As the drives are dual-controller, two NVMe
devices -- of half its physical size -- are exposed by the system for each physical device.
In order to expose a single NVMe consisting of
its two controllers, or to unify several NVMe devices, Intel Rapid Storage
Technology \cite{intel_rst} (RST) is used. RST composes a RAID0 of the
controllers which becomes exposed over fabric as a single NVMe card.  Mellanox OFED
4.0-2.0.0.1 drivers were used for the InfiniBand HCA adapters. The drivers
included modules for NVMe over fabrics as well, both the target and the
client. Kernel 4.8.0-39 was used under Ubuntu server 16.10 operating system
in all nodes.

We use SMUFIN on its merge stage, as explained in section~\ref{s_smufin}.  In
the following evaluations each SMUFIN instance reads and processes a sample
DNA input (+300GB) from a NFS shared storage, while the shared NVMe devices
are used as memory extension for temporary data and final output. SMUFIN has
been implemented to maximize sequential writes to the devices, and this
behavior has been verified by analyzing its access pattern.
A block trace sample of requested blocks to the device was generated using
Linux's \textit{blktrace}, and the trace was then fed to the
algorithm provided by \cite{accesspattern} to calculate the percentage of
sequential write accesses.  This method identified 88\% of sequential writes
after adapting the algorithm to consider accesses in which the final address matched
the initial address of many immediately following requests, thus accounting for file
appends.

\subsection{Direct-Attached Storage vs NVMe over Fabrics}

\begin{figure*}[t]
\vspace{-0.5em}
	\centering
	\includegraphics[width=0.85\textwidth]{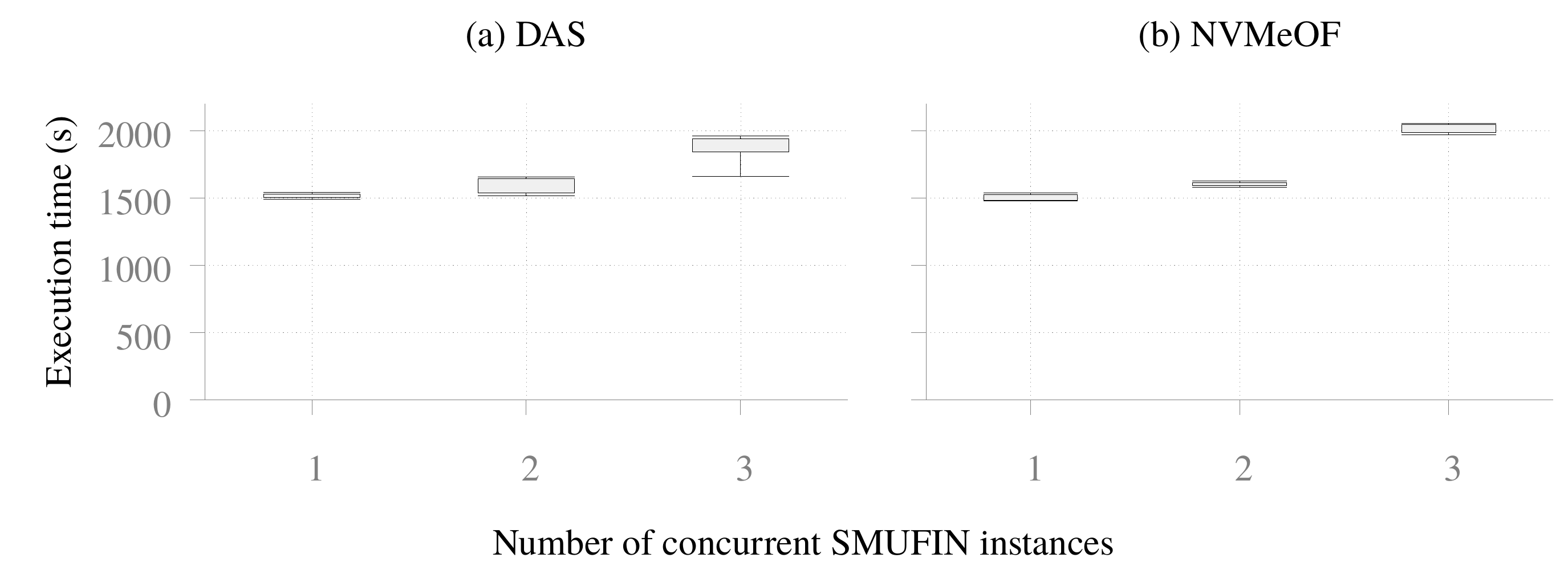}
    \caption{Boxplot of execution time of Direct-Attached Storage (DAS) and NVMeOF when running 1x, 2x and 3x SMUFIN instances on the same node}
	\label{fig:das_nvmeof}
\vspace{-0.5em}
\end{figure*}

The performance of NVMeOF has been studied in the
literature~\cite{lowoverhead_nvmeof}, and found not to show any significant
degradation when compared to \textit{local} directly-attached storage (DAS).
Additionally, in this section we perform our own experiments running up to 3
instances of SMUFIN in the same node:
against a directly-attached NVMe device and against NVMeOF.
Each instance processes the same dataset, generating
$\approx$150GB, with an average use of bandwidth of 477MB/s per SMUFIN
instance. The NVMe device is capable of handling 2GB/s bandwidth under
sequential write pattern, as is the SMUFIN scenario.
Figure~\ref{fig:das_nvmeof} shows average execution time and deviation after
repeating the executions six times. As it can be observed, when running one
and two instances on local storage (~\ref{fig:das_nvmeof}a) there is no
performance degradation when disaggregating NVMe over fabrics
(~\ref{fig:das_nvmeof}b). However, when running three concurrent instances
there is a significant degradation of 6\% when using NVMeOF.

\begin{figure*}[t]
	\centering
	\includegraphics[width=0.85\textwidth]{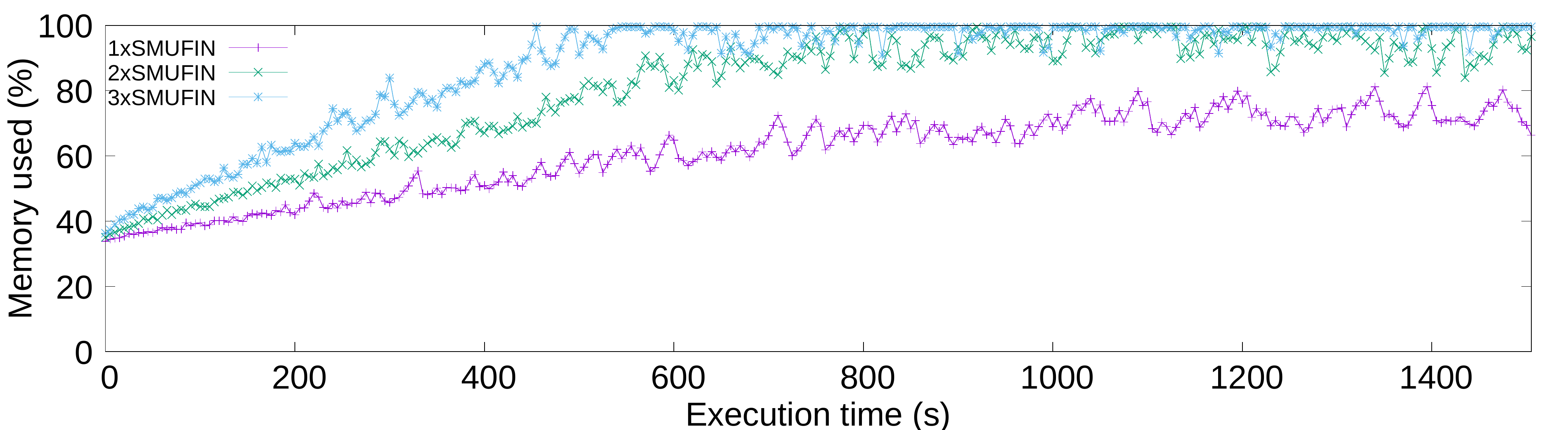}
	\caption{Memory usage on 1x,2x,3xSMUFIN scenarios using Direct-Attached NVMe on a period of 1500 seconds}
	\label{fig:das_memory}
\vspace{-1.5em}
\end{figure*}

On the other hand there is a certain performance degradation
scaling up to three instances in both scenarios. Analyzing this behavior, up
to two instances, the host's memory can handle all the intermediate data
generated by SMUFIN and the NVMe becomes only used to output final data.
However, with three instances the memory becomes a bottleneck and intermediate
data not fitting in memory gets flushed to the NVMe device more frequently. Is
in this scenario when degradation is observed and performance comparison
against NVMeOF is worse. Figure \ref{fig:das_memory} depicts memory usage on
the three scenarios (1, 2 and 3 SMUFIN on the same node, directly-attached)
over a period of 1500 seconds, evidencing the memory bottleneck.

\subsection{Resource Sharing And Composability}
\label{sec:wlscalability}

When multiple workloads share resources and hence compete for its usage, their
execution time compared to a dedicated execution in isolation degrades when a threshold
is reached, as shown in previous section. In this section we explore if degradation
still occurs when running up to six concurrent instances, all of them using
partitions from the same set of NVMe devices and running on separate nodes to avoid the aforementioned interferences.

\begin{figure}
\vspace{-1.5em}
\centering
\includegraphics[width=0.90\textwidth]{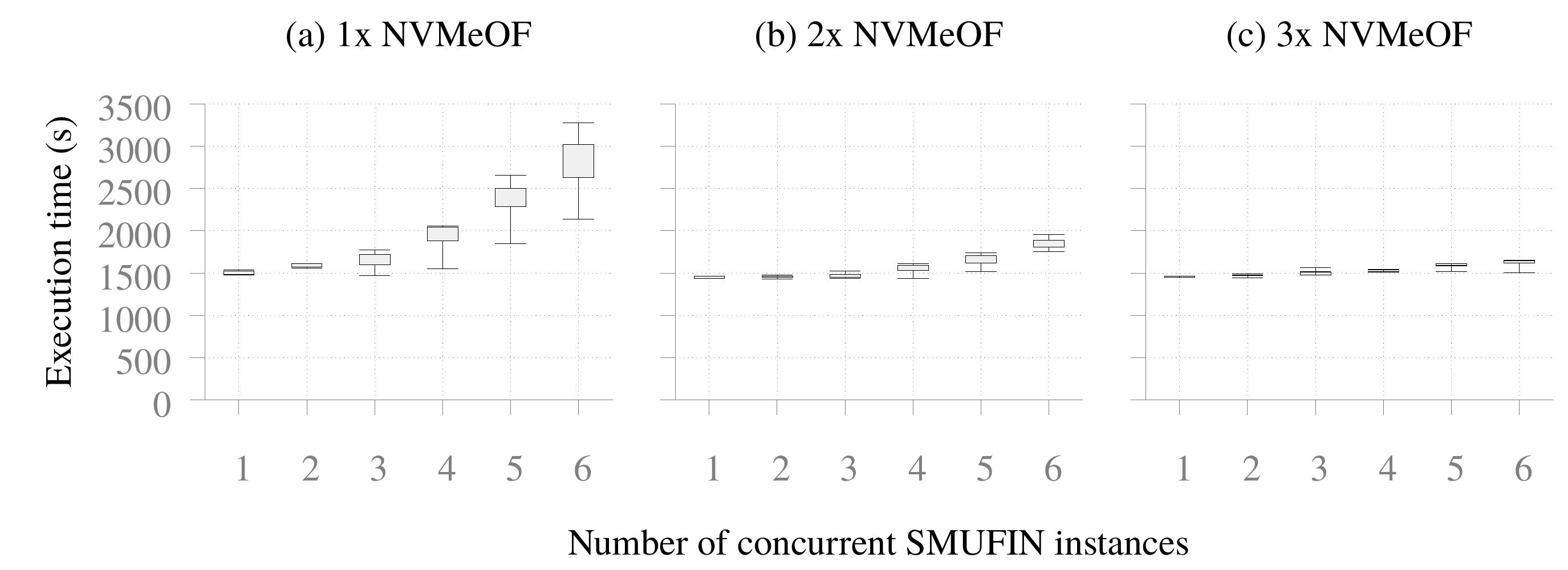}
\caption{Boxplots showing how execution time evolves when running multiple
    SMUFIN instances: sharing a single device (1xNVMe), or sharing on composed
    nodes (2xNVMe, 3xNVMe)}
\label{fig:smufin_nvme_exetime}
\vspace{-1.5em}
\end{figure}

Figure ~\ref{fig:smufin_nvme_exetime}a 
represents the box plot of individual execution times under different configurations, 
along with its quartiles, median, and standard deviation. In (a) only one NVMe SSD
is used. It can be observed running three instances separately against a single device 
do not degrade as significantly as running under the same node. However, performance degradation
is still experienced when certain resource sharing threshold is reached.

When disaggregating NVMe over fabrics we can benefit of composing several NVMe devices and expose
them as a single one. Under composition, profiling data
shows that the Intel driver balances the bandwidth evenly through all composed
devices. It is also observed that provided bandwidth scales linearly with the number
of devices, hence under 2 and 3-compositions 4GB/s and 6GB/s of sequential write speed can be reached 
respectively (each individual drive provides 2GB/s). 
Through composition, performance degradation can be mitigated.
Compositions of two and three NVMe SSD exposed as a single target to clients
increases the bare-performance, as a composition multiplies the total available
bandwidth. The evolution of execution time respect composition level is presented in 
Figures \ref{fig:smufin_nvme_exetime}b and
\ref{fig:smufin_nvme_exetime}c. In the 2-composition scenario, up to 3
sharing workloads obtain the same performance as if running alone in a single
NVMe. The level of concurrency can be increased without introducing
significant degradation using a composition of 3 NVMe, being able to have six
sharing workloads with a similar performance as when running alone in a single
device. Thus, workloads indeed benefit of resource composition. However, in all scenarios 
performance degradation still occurs on reaching a certain
threshold, larger as more devices are used. Under 2-NVMe compositions it is at four workloads, whereas
on the 3-composition the tendency is observed at six instances threshold. 

\subsection{Bandwidth}
\label{sec:bwdistribution}

We observed performance degradation when a certain sharing ratio of resources is reached. Despite composition increases this threshold, 
degradation still occurs regardless of composition. As the memory bottleneck was removed and cannot be found on the network bandwidth, we analyze
the target NVMe bandwidth.

\begin{figure}
\vspace{-1.0em}
\centering
\subfloat[1x NVMe]{\includegraphics[width=0.47\textwidth]{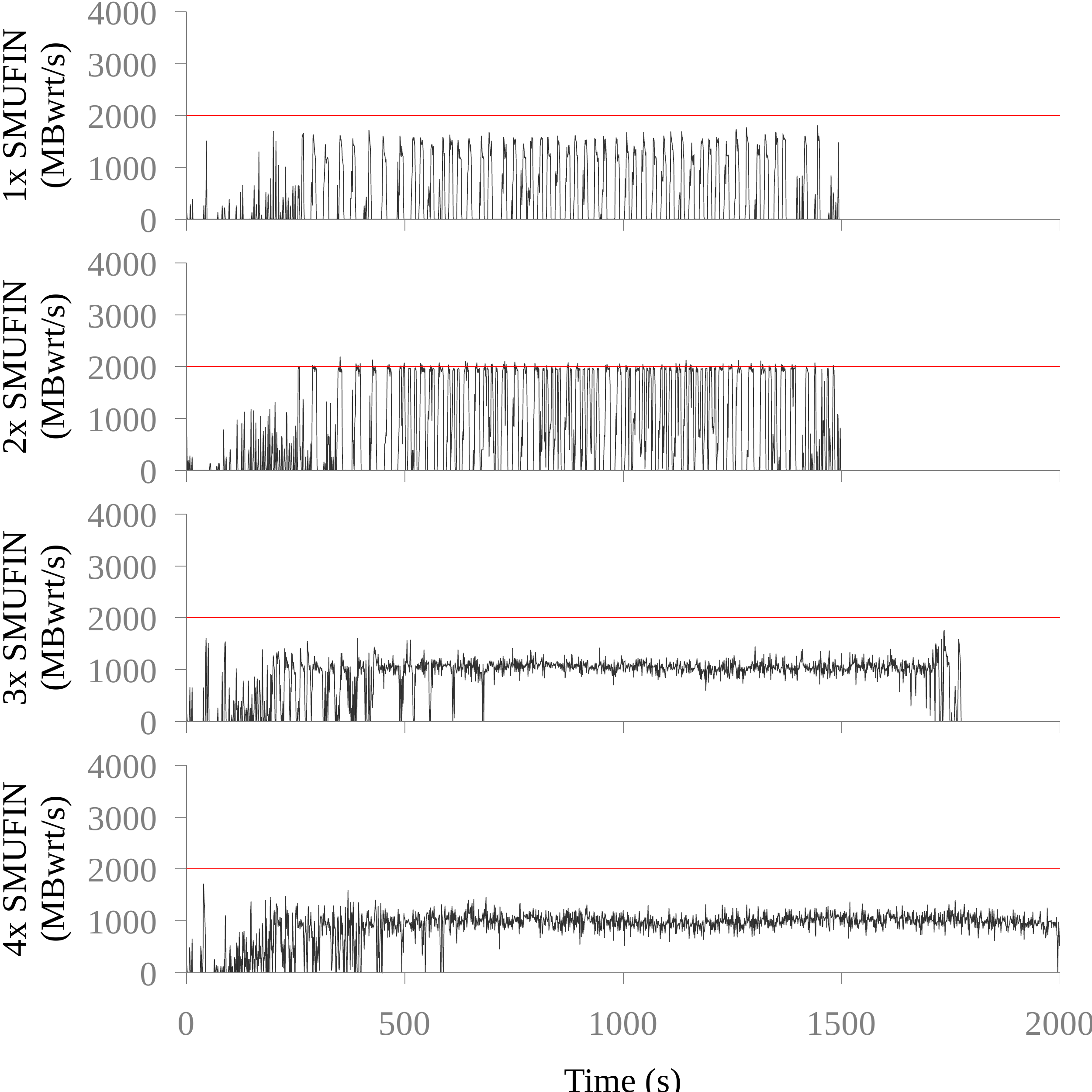}\label{fig:bw_usage1xnvme}}\qquad
\subfloat[2x NVMe]{\includegraphics[width=0.47\textwidth]{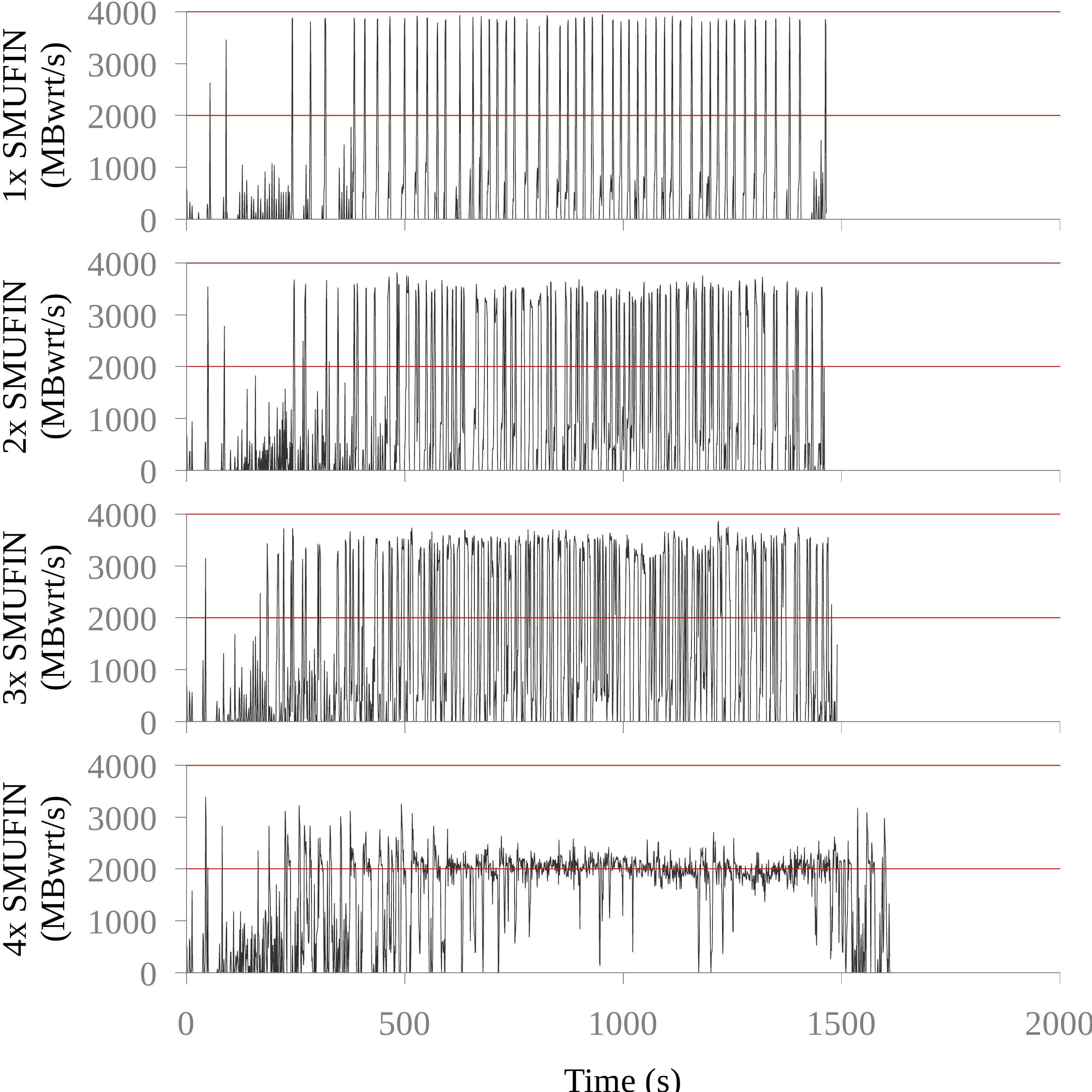}\label{fig:bw_usage2xnvme}}\qquad
\caption{Bandwidth measured from the NVMe pool server for 1x, 2x, 3x and 4x instances of SMUFIN}
\label{fig:bw}
\vspace{-1.0em}
\end{figure}

Figures~\ref{fig:bw_usage1xnvme} and~\ref{fig:bw_usage2xnvme}
show the NVMe bandwidth over time for experiments running up to four concurrent SMUFIN instances in the single-resource and the 2-composed resource configuration.
The solid horizontal lines indicate the maximum bandwidth for sequential write
that the resources can provide (2GB/s in single-resource configuration, 4GB/s for the
composed scenario).

From the figures it can be appreciated, on one hand, that resource composition
scales linearly, doubling the maximum available bandwidth of a single resource.
In both scenarios, two important characteristics can be noticed as more concurrent instances are included
in the experiment: (1) the bandwidth observed from the NVMe perspective is steadier; 
(2) the bandwidth that the NVMe device is capable of delivering is
reduced as more concurrent instances are added. Running a single instance, the
full bandwidth of the combined NVMe can be used with bursts at the maximum 4 GB/s.
However as more concurrent executions are added these bursts make
use of less bandwidth until reaching saturation levels, decreasing
significantly.

\section{Towards Efficient Orchestration of Shared and Composed Resources}
\label{s_orchestration}

Previous sections have shown how NVMe disaggregation provides new ways to use
resources through resource sharing and composition. However, its behavior is
not obvious a priori: heavy resource sharing may have a negative impact on
performance, whereas composition may help increase sharing ratios without
degradation. Therefore, deciding whether to compose a resource or to share it
among many workloads is not trivial decision. With the help of workload
characterization, platform orchestrator will be able to make more informed
and smarter decisions.

\begin{figure}[t]
\vspace{-1.0em}
	\centering
	\includegraphics[width=0.95\textwidth]{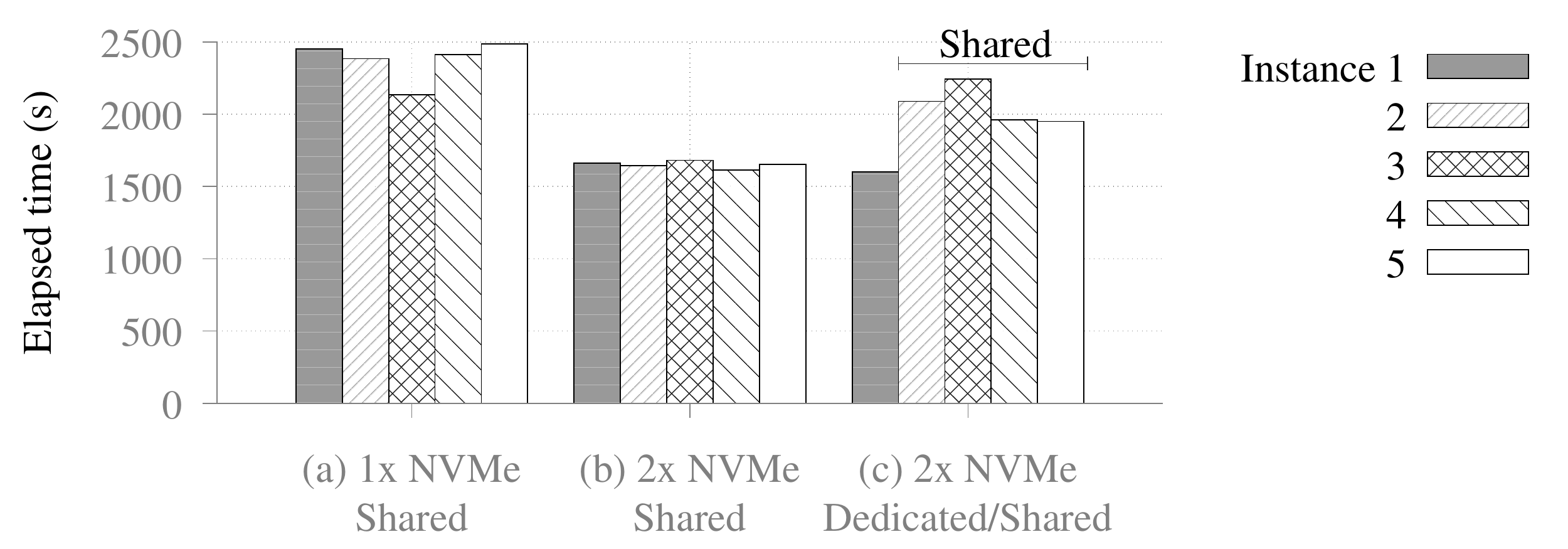}
	\caption{Execution time of 5 SMUFIN instances under different scenarios}
	\label{fig:schedule_example}
\vspace{-1.0em}
\end{figure}

In Figure~\ref{fig:schedule_example} we present different orchestration
policies that could be managed with our data. The figure shows our cluster
running five concurrent instances of SMUFIN, and three different resource
allocation strategies for the instances: a) sharing a single device, b)
sharing two NVMe devices, and c) one instance-dedicated device and the
remaining four instances on a shared NVM device.  This example was run under
the same setup as in section \ref{sec:evaluation}.  When the SMUFIN instances
use two composed devices (b) it leads to faster executions times than using a
single device (a). However, when using a dedicated device to run a single
instance and a shared device to run the remaining four (c), the
dedicated-device does not grant that instance an improved performance compared
to a fully shared scenario using both devices (b).  Intuitively it might be
believed that just sharing all the resources under composition is the obvious
winning strategy.  However this approach does not consider arriving workloads
might have a time requirement for completion, and upon arrival of those
workloads, if the resources are fully occupied serving others the orchestrator
will be unable to meet the requirement. Other concerns might be power consumption
or total cost of ownership (as more resources, more expensive it becomes to run).
Therefore, the strategy to follow must
consider the trade-off between execution time and requirements of current and
incoming workloads to maximize the granted quality of service, which in the
case of genomics might be critical.  The work on those policies is out of the
scope of this paper and left as future work.

\section{Related work}
\label{sec:related}


Genomics workloads and pipelines in general are a good fit for disaggregation,
but prior to this paper applications haven't explored its large-scale
explotation. A number of different approaches to parallelize whole genome
analysis in HPC systems have been proposed in the
literature~\cite{btu071},~\cite{kawalia2015}, and~\cite{li2009snp}, but these
tend simply adapt existing algorithms without considering or taking complete
advantage of next generation computing platforms.



%

Resource disaggregation is being increasingly studied in the literature.
In~\cite{netreq_resource_disagg}, the authors examine the network requirements
for disaggregating resources at rack- and data-center levels. Minimum
requirements are measured in terms of network bandwidth and latency. 
Those requirements must be such so that a given set of applications doesn't
experiment any performance degradation when disaggregation memory or other
resources over the fabric.
\cite{asplos17_nvmf} implements NVMe disaggregation, but unlike the work presented in this
paper, the authors focus on a custom software layer to expose devices instead
of using the NVMeOF standard.
On the other hand, \cite{fpga_disagg_cloudcom16} evaluates the impact of FPGA
disaggregation.
In terms of Software-Defined Infrastructures, Intel Rack Scale~\cite{intel_rsa} is a prototype
system that allows dynamic composition of nodes. It fully disaggregates resources in pools, such as
CPU, storage, memory, FPGA, GPU, etc. 
Facebook has engaged with Intel developing its own prototype, the Facebook Disaggregated
Rack~\cite{facebook_rack}.

\section{Conclusions}
\label{sec:conclusions}

This paper evaluates resource sharing and composition benefits for NVM-centric workloads in the context of disaggregated datacenters. 
This work takes SMUFIN, a real production workload in the field of Computational Genomics, leveraging remote NVMe devices as memory extension.
This paper presents a comprehensive characterization of SMUFIN's resource consumption patterns. It is shown NVMe is utilized in a sequential write pattern.
A performance comparison between directly-attached NVMe and NVMeOF is then presented and shown that as long the system's memory is capable of handling SMUFIN
instances there is no degradation. To increase concurrency disaggregating over fabrics allows to share the same resource across multiple nodes running instances, as well as the possibility of composition. Thus, through disaggregation we are able to handle more concurrent SMUFIN instances without individual degradation.
On the other hand, reaching the resources' sharing ratio limit significantly degrades performance as the utilization of the available bandwidth diminishes, never reaching its maximum. Thus the NVMe becomes the bottleneck.

Finally the paper briefly explains how the results of this characterization could be used to implement data-center scheduling policies in order to maximize the efficiency in terms of Quality of Service. Quality of Service could be understood in terms of execution time, so all workloads should be completing its executions within a certain requested time-frame. The work on those policies is left as future work.

\vspace{-1em}
\subsubsection*{Acknowledgment}

This work is partially supported by the European Research Council (ERC) under
the EU Horizon 2020 programme (GA 639595), the Spanish Ministry of Economy,
Industry and Competitivity (TIN2015-65316-P) and the Generalitat de Catalunya
(2014-SGR-1051).

\bibliographystyle{splncs03}
\bibliography{references}

\end{document}